# Laser-diode-heated floating zone (LDFZ) method appropriate to crystal growth of incongruently melting materials

Toshimitsu Ito[1,*], Tomoharu Ushiyama[1], Yuji Yanagisawa[1], Yasuhide Tomioka[1], Isamu Shindo[2], and Atsushi Yanase[3]

[1] National Institute of Advanced Industrial Science and Technology (AIST), Tsukuba, Ibaraki 305-8562, Japan

[2] Crystal Systems Corporation, Hokuto, Yamanashi 408-0041, Japan

[3] Miyachi Corporation, Noda, Chiba 278-0016, Japan

[*] Corresponding author. E-mail: t.ito@aist.go.jp, Tel.: +81-29-861-5112, Fax: +81-29-861-2586

ABSTRACT

We have developed the laser-diode-heated floating zone (LDFZ) method, in order to improve the broad and inhomogeneous light focusing in the conventional lamp-heated floating zone method, which often causes difficulties in the crystal growth especially for the incongruently melting materials. We have simulated the light focusing properties of the LDFZ method to make irradiated light homogeneous and restricted mostly to the molten zone. We have designed and assembled an LDFZ furnace, and have demonstrated how it works through actual crystal growth. The method is applicable to various kinds of materials, and enables stable and reproducible crystal growth even for the incongruently melting materials. We have succeeded in the crystal growth of representative incongruently melting materials such as $BiFeO_3$ and $(La,Ba)_2CuO_4$, which are difficult to grow by the conventional method. Tolerance to the decentering of the sample and highly efficient heating are also established in the LDFZ method.

Highlights

► The laser-diode-heated floating zone method have been developed. ► The irradiated light is homogeneous and restricted mostly to the molten zone. ► The temperature gradient at the interface between the liquid and the solid is steep. ► Stable and reproducible crystal growth is realized even for the incongruently melting materials. ► Tolerance to the decentering of the sample and highly efficient heating are established.

1. Introduction

The conventional lamp-heated floating zone (FZ) method has become widespread through the research on the high-$T_c$ superconductors [1], and has been a powerful tool for the researchers in the field of the correlated electron systems [2]. It has been noticed that for some of the incongruently melting materials, the crystal growth by this method with the travelling solvent floating zone (TSFZ) technique is difficult or virtually impossible, even if the growth should be possible when judged from their phase diagrams. It is because the light focus is broad and the temperature gradient at the interface between the solid and the liquid is gentle, which makes the melt seriously attack the feed rod and spill over the crystal and eventually makes the growth unstable. In addition, inhomogeneous heating along the rotational direction degrades the quality of the crystal. In order to solve these problems, we propose and demonstrate a method that can confine the irradiated lights on the molten zone homogeneously by use of laser beams, whose focusing can be designed as intended.

2. Problems in the conventional lamp-heated FZ method

The problems in the focusing properties along the axial and the rotational directions in the conventional lamp-heated FZ method with ellipsoidal mirrors are discussed in the following subsections. Here we consider a furnace with two ellipsoidal mirrors. For a furnace with four ellipsoidal mirrors, the problems are less prominent but still remains to some extent.

2.1. Problems in the focusing properties along the axial direction

Assuming a point light source for the conventional lamp-heated FZ method, one can trace various light paths in a vertical cross section, as shown in the left half of Fig. 1 (a). Since the sample has finite diameter, the light is irradiated in a wide range along the axial direction. The light intensity of the irradiated light on the sample surface is schematically shown in a color scale. As shown in the right half of Fig. 1 (a), actual filament has a finite size, which broadens the irradiated light more. Such broad focusing makes the temperature gradient along the axial direction gentle, and then makes the melt seriously attack the feed rod and spill over the crystal.

There is a report on a trial to make light focusing area narrower by screening the light paths that reflect at the top or the bottom part of the mirrors, or by using flat filaments [3]. The trial is effective to some extent. In the former case, however, the efficiency in the use of the light is reduced and the achievable temperature becomes lower.

When the sample position is decentered due to the misalignment of the sample or the contact of the feed rod and the crystal, as shown in Fig. 1 (b), the distribution of the irradiated light intensity as well as that of the temperature is modified. The irradiated light on the sample surface far from the rotation axis (the left-hand side surface in the figure) is focused more broadly and its temperature is lowered.

2.2. Problems in the focusing properties along the rotational direction

Assuming point light sources, one can trace various light paths in a horizontal cross section, as shown in Fig. 1 (c). The light intensity of the sample surface faced to the light sources is higher than that of the surface away from it. By the rotation of the crystal during the growth, its surface temperature oscillates due to this focusing property, which causes undesirable instantaneous solidification of the melt. By the rotation of the sample and the stirring of the melt, inhomogeneity in the temperature distribution becomes less but remains to some extent.

When the sample position is decentered, as shown in Fig. 1 (d), the distribution of the irradiated light intensity as well as that of the temperature is modified. The irradiated light on the sample surface far from the rotation axis (the left-hand side surface in the figure) is focused broadly and its temperature is lowered.

3. Solution of the problems by the LDFZ method

Due to the limitation in the focusing properties for the lamp-heated FZ method as described above, we propose the LDFZ method to improve the focusing. By use of laser beams, as shown in Fig. 1 (e), the light can be concentrated on the molten zone. When equivalent lasers and optics are arranged at the equal intervals on a circle whose center is at the sample position and emit laser beams to the sample, as shown in Fig. 1 (f), homogeneous irradiation along the rotational direction is expected. A schematic perspective picture of the LDFZ method is shown in Fig. 2.

Laser diode (LD) or semiconductor laser is most appropriate to this purpose, since its available laser power is increased and its price is lowered in recent years due to the striking technological progress. We consider a parallel laser beam that has a rectangular cross section with homogeneous light intensity inside. Such a beam is realized approximately by use of a homogenizer such as a light pipe.

At the edge of the beams in Fig. 1 (e), the irradiation intensity along the axial direction changes discontinuously and the temperature gradient is maximal. In order to make the interface between the solid and the liquid flat and smooth, the interface should be located near the edge of the beam where the temperature gradient is maximal, which can be realized by the adjustment of the laser power. The decentering of the sample in Fig. 1 (e) does not affect the light intensity distribution along the axial direction by use of the parallel laser beam.

We can make the irradiated light intensity along the rotational direction homogeneous, as is shown in Fig. 1 (f), which will be discussed in details in the next section. There is no influence on heating by the decentering of the sample, when the whole of the decentered sample is within all of the laser beams, namely, inside the polygon formed by the overlapped beams, as shown by dark red area in Fig. 1 (f).

4. Simulation of the light intensity distribution on the sample surface by the LDFZ method

It is expected that increased number of LDs improve the homogeneity of the light intensity along the rotational direction. To design an LDFZ furnace with appropriate number of LDs, we simulate the light intensity distribution on the sample surface as a function of the number of LDs ($N$). Assuming a cylindrical sample, one-dimensional distribution along the rotational direction is simulated. The results for $N$ = 3, 4, 5, 6, 7, and 8 are shown in Fig. 3. The irradiated light intensity from each LD (red dashed line) is represented by a cosine of the beam incident angle for the surface faced to the LD, and is zero for the surface on the opposite side. The total intensity from all the LDs (blue solid line) is obtained by the summation of the intensity from each LD. The total intensity becomes smoothed by the summation. An angle at a valley bottom in the total intensity distribution corresponds to that where the incident angle is 90° for one of the LDs.

We define the ratio of minimum to maximum in the total intensity as homogeneity, which is plotted as a function of N in Fig. 4. The overall feature is that the homogeneity increases and approaches to 100% with increasing N. A close watching reveals that the homogeneity oscillates and is better for an odd number N compared to that for the adjacent even numbers N-1 and N+1. The homogeneity for an odd number N equals to that for the even number 2N that is twice of the former. To achieve certain homogeneity, a half number of LDs are sufficient in the case of the odd number, compared to the case of the even number. In the series limited to odd or even numbers, the homogeneity increases with increasing N.

We notice that the number of valleys for odd N is more than that for adjacent even numbers N-1 or N+1, as shown in Fig. 3. In the case of even numbers, a pair of laser beams point to antiparallel directions and have two common valleys, indicating that total number of valleys equals to N. In the case of odd numbers, laser beams do not point to antiparallel directions and each laser beam has two valleys independently, indicating that total number of valleys equals to 2N. Such increase in the number of valleys for odd N makes the intensity pattern fine and smooth, and suppresses the decrease of intensity at the valley bottoms. It is worth to note that the intensity distribution for N = 3 coincides completely with that for N = 6 in Fig. 3. In general, the intensity distribution for the odd number N coincides with that for the even number 2N.

From the discussion so far, odd number of LDs are effective to homogenize the irradiated light along the rotational direction. In addition, in the case of even number of LDs, the laser beam emitted from a LD may enter and damage the LD at the symmetric position, which also favors odd number of LDs.

5. Assembling of the LDFZ furnace

Based on the simulations, we have adopted 7 LDs, which ensures the homogeneity of 97.5 %. We assembled a furnace shown in Fig. 5. The wavelength of LDs is 975 nm. Maximum power of each LD is 50 W and the maximum total power is 350 W. In order to keep the emitted laser powers of all the LD identical, the power of each LD is calibrated and individual power source is used to control each LD. In order to stabilize the laser power, Peltier module is used and the temperature of the LDs is kept constant within ± 0.01 K. By this system, ruby (melting point of ~2000 °C) can be melted.

6. Examples of crystal growth by the LDFZ method

It is well known that the crystal growth of incongruently melting $BiFeO_3$ by the conventional lamp-heated FZ method is difficult. (There are a few reports for the growth by the conventional FZ method [4].) Our trial by the conventional method is shown in Fig. 6 (a). Due to the gentle temperature gradient along the axial direction, the melt attacks the feed rod, which induces inhomogeneous melting and makes the growth unstable. In addition, the melt often spills over the grown crystal. There is no clear border between the molten zone and the feed rod or the grown boule. In contrast, in the case of the LDFZ method, as shown in Fig. 6 (b), the temperature gradient is steep, the interface between the liquid and the solid is clear and flat, the melt scarcely attacks the feed rod and scarcely spills over the grown crystal, and therefore long-term stable growth is realized. A grown crystal is shown in Fig. 6 (c). The details of the growth is described in ref. [5]

It is also known that the crystal growth of incongruently melting $La_{2-x}Ba_xCuO_4$ by the conventional lamp-heated FZ method is difficult. (There are a few reports for the growth by the conventional method [6-8].) By the LDFZ method, as shown in Fig. 7 (a), stable crystal growth is realized and crystals with $x > 1/8$ are obtained. A grown crystal with a flat facet is shown in Fig. 7 (b). The details will be described elsewhere.

7. The characteristic features of the LDFZ method

The features of the LDFZ method are summarized below, revealed by the actual crystal growth.

7.1. Focusing properties along the axial direction by the LDFZ method

Homogeneous irradiation of laser beam, whose cross section mostly covers the molten zone, is realized by use of a light pipe. Typically, the vertical dimension of the molten zone equals to that of the beam multiplied by 0.8 – 0.9. It is a great advantage for the beam size to be easily designed to fit the desired size of the molten zone. The sharp cutoff at the edge of the beam prevents the melt from attacking the feed rod and from spilling over the crystal, and makes the interface between the solid and the liquid flat and clear. Such features are useful to stabilize and reproduce the crystal growth reasonably well. Precise temperature control of the LDs by the Peltier module contributes to the stable crystal growth as well, and does not require one to adjust the operation currents of the LDs for a period longer than a week under the optimized conditions.

The temperature gradient near the interface between the solid and liquid for $BiFeO_3$ can be estimated here. It is likely to identify the interface between the molten zone and the feed rod as being at the eutectic point (933 °C [9]), and the interface between the feed rod absorbed with the melt (the thin glittering part just above the molten zone in Fig. 6 (b)) and the feed rod without the melt as being at the peritectic point (777 °C [9]). Since the distance between these two interfaces for the LDFZ method is only ~1 mm from Fig. 6 (b), the average temperature gradient of this region is estimated to be ~150 °C/mm. In the case of the conventional lamp-heated FZ method, the corresponding distance is ~5 mm from Fig. 6 (a), and the average temperature gradient of this region is estimated to be ~30 °C/mm.

The LDFZ method satisfies efficient local heating. In the case of ruby, the total input electric power into LDs is 630 W and the total output laser power is 280 W for the LDFZ method, whereas the total input electric power is 2100 W and the total output light power is estimated to be 1900 W under the assumption that the efficiency of the halogen lamp is 90 % for the lamp-heated FZ method. In the case of BiFeO$_3$, the total input electric power into LDs is 180 W and the total output laser power is 30 W for the LDFZ method, whereas the total input electric power is 190 W and the total output light power is estimated to be 170 W under the same assumption of the efficiency described above for the lamp-heated FZ method. Therefore, the output light power required for the lamp-heated FZ method is about 6 times as much as that for the LDFZ method. It suggests that most of the light power is not used to heat the molten zone directly in the case of the lamp-heated FZ method.

7.2. Focusing properties along the rotational direction by the LDFZ method

In the case of the odd number of LDs, the intensity of the irradiated light becomes homogeneous effectively, as shown by the simulation. For the assembled furnace with N = 7, the interface between the solid and the liquid is flat, as shown in Fig. 6 (b), suggesting that the homogeneity of the heating power distribution as well as the temperature distribution along the rotational direction is realized. It is expected that such homogeneous heating would reduce defects in the grown crystals.

By making the horizontal width of the homogeneous beam larger than the diameter of the sample, the crystal growth is stable even if the misalignment of the sample or the displacement of the sample by the contact of the feed rod and the crystal occurs, displaying that the intensity of the irradiated light is unaffected. From all of these features stable crystal growth is realized.

8. Conclusions

We have developed the LDFZ method optimized to the crystal growth of the incongruently melting materials. A multiple number of equivalent LDs and optics are arranged in equal intervals on a circle whose center is at the sample position. In the optics, light pipes are used to make the cross section of each laser beam rectangular and the distribution of light intensity homogeneous. The laser beams emitted from them are irradiated on the sample, which realizes homogeneous heating along the rotational direction. Odd number of laser beams are effective to homogenize irradiation. Along the axial direction, the dimension of each beam can be designed to nearly equal to the desirable vertical length of the molten zone. Such a design makes the temperature gradient at the interface between the liquid and the solid steep and prevents the melt from attacking the feed rod and from spilling over the crystal. These features enable stable crystal growth. The horizontal width of each beam is designed to be larger than the diameter of the sample, which makes the irradiated light intensity unchanged if the misalignment or the decentering of the sample occurs. The precise temperature control of the LDs by the Peltier module ensures the stability of the irradiated light intensity. Thus we have overcome the problems in the lamp-heated FZ method.

ACKNOWLEDGEMENTS

This work was partly supported by JSPS Grants-in-Aid for Scientific Research (Grant Number 22560018) and the Funding Program for World-Leading Innovative R&D on Science and Technology (FIRST Program), Japan.

FIGURE CAPTIONS

Figure 1

Schematic drawings of the light paths and the distribution of the irradiated light intensity on the sample surface for the conventional lamp-heated FZ method and the LDFZ method. To make the effects displayed clearly, the diameter of the sample is drawn larger than the actual size. (a) Vertical cross section of the conventional lamp-heated FZ method. The left half is for the point light source and the right one is for the light source with finite size such as a filament. (b) Vertical cross section of the conventional lamp-heated FZ method for the point light source, for the case that the sample position is decentered. (c) Horizontal cross section of the conventional lamp-heated FZ method for the point light source. (d) Horizontal cross section of the conventional lamp-heated FZ method for the point light source, for the case that the sample position is decentered. (e) Vertical cross section of the LDFZ method. (f) Horizontal cross section of the LDFZ method.

Figure 2

A schematic perspective drawing of the LDFZ method. Homogeneous laser beams whose cross section is rectangular are irradiated.

Figure 3

The simulation of one-dimensional distribution of irradiated light intensity along the rotational direction on a cylindrical sample. Homogeneous laser beams, as shown schematically in Figs. 1 (e), 1 (f) and 2, with N = 3, 4, 5, 6, 7, and 8 are irradiated. The red dashed line is for each LD and the blue solid line is for the total of all the LDs. The intensities are plotted in arbitrary unit.

Figure 4

The homogeneity, defined as the ratio of the minimal to the maximal intensities along the rotational direction, as a function of N.

Figure 5

Photographs of the assembled LDFZ furnace. (a) The sample chamber and the heating elements. (b) The whole of the furnace.

Figure 6

The crystal growth of $BiFeO_3$ (a) by the conventional lamp-heated FZ method and (b) by the LDFZ method. (c) A grown crystal by the LDFZ method.

Figure 7

(a) The crystal growth and (b) a grown crystal of $(La,Ba)_2CuO_4$ by the LDFZ method.

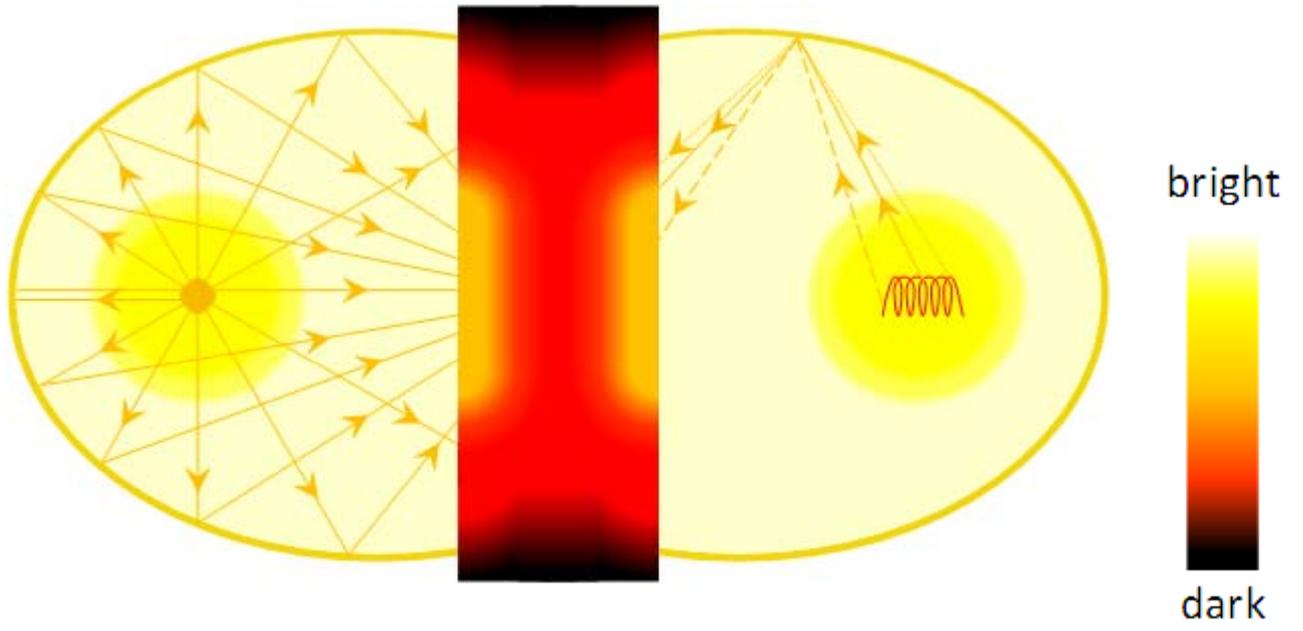
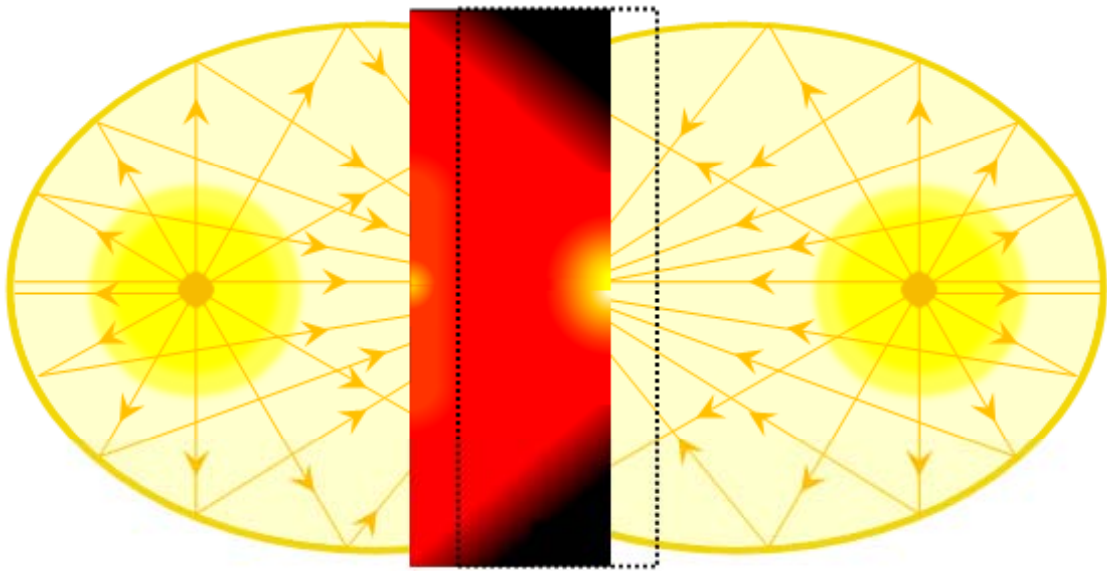

Fig. 1

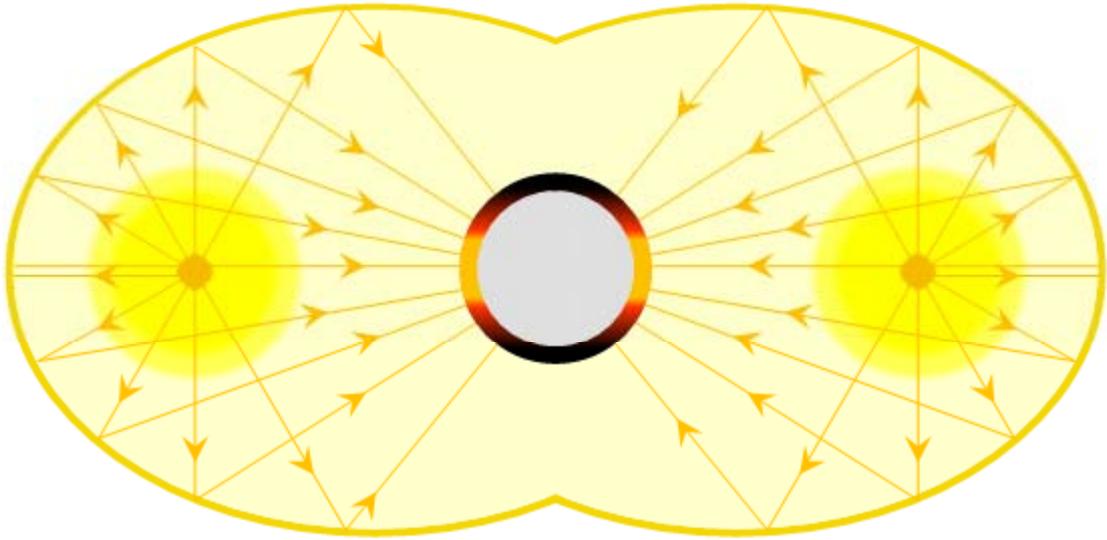

(c)

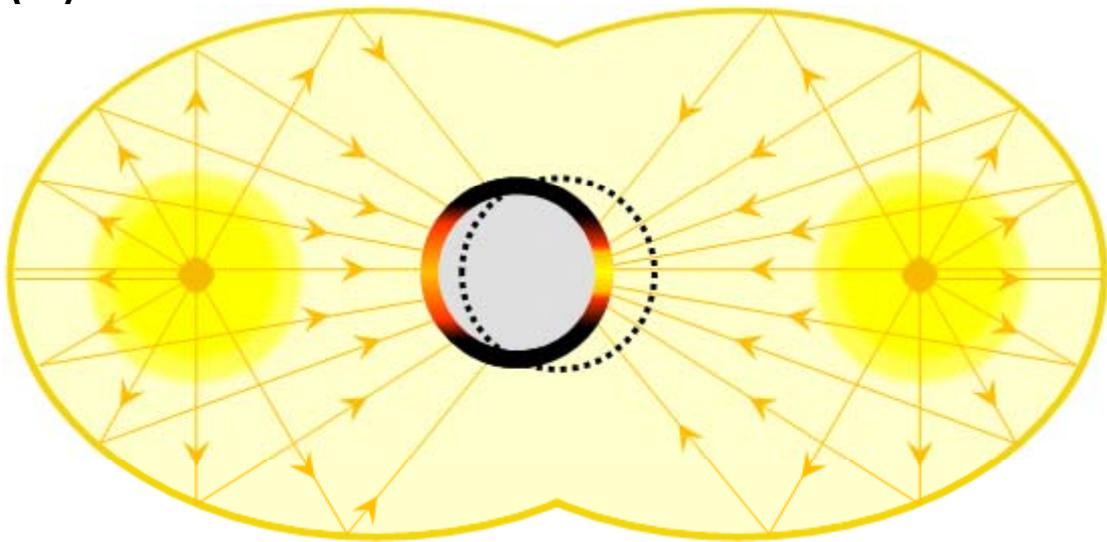

(d)

Fig. 1

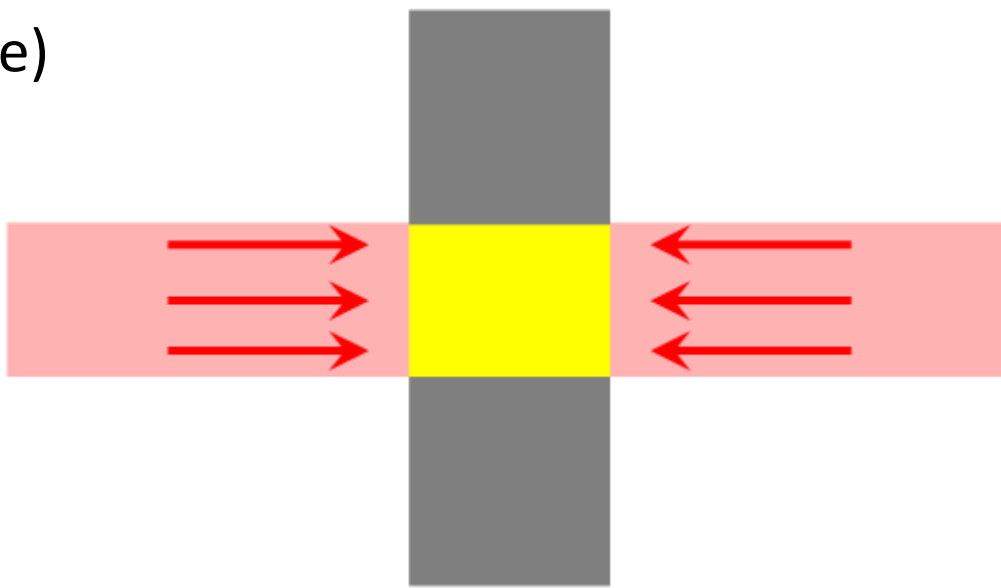

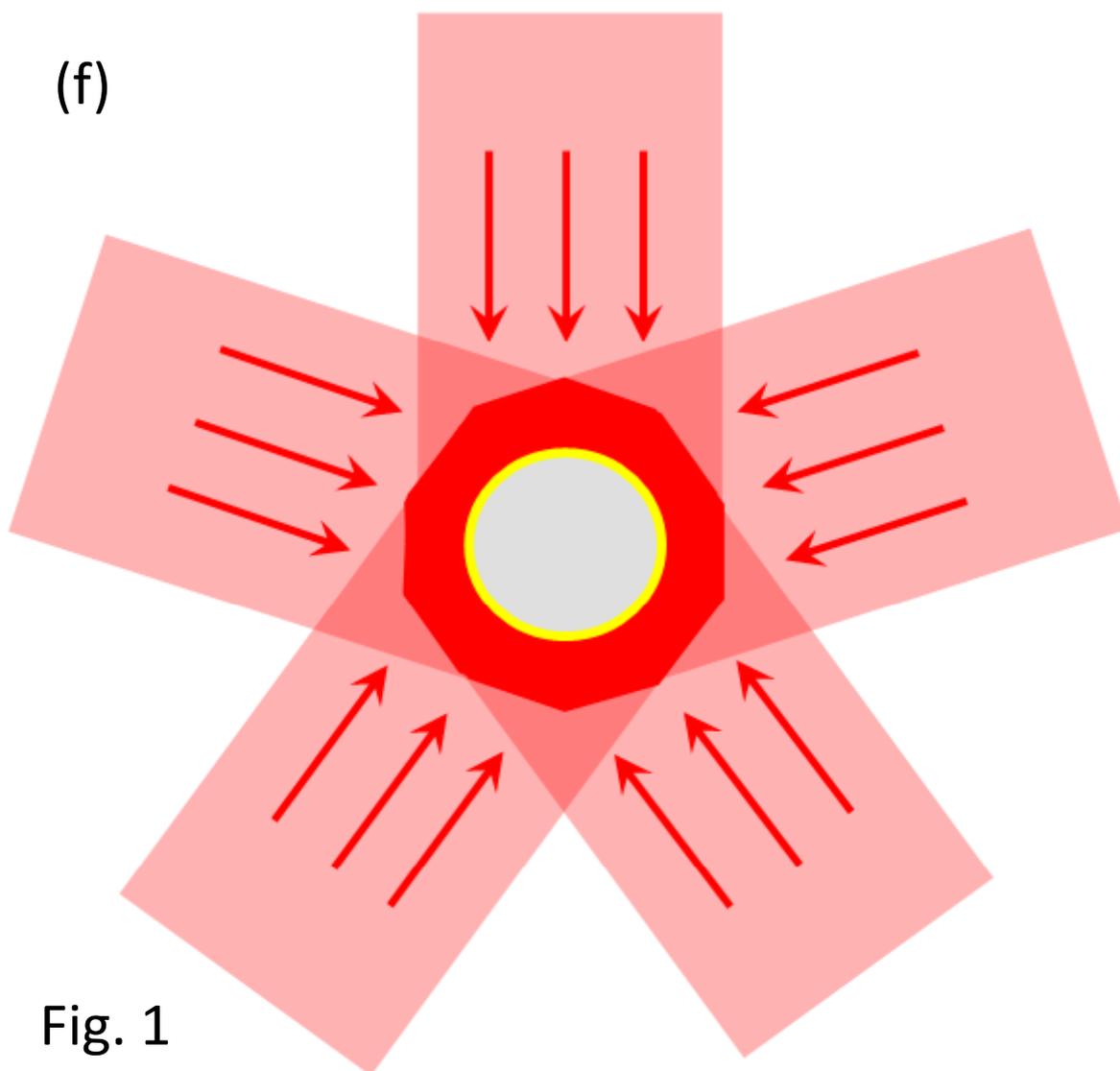

Fig. 1

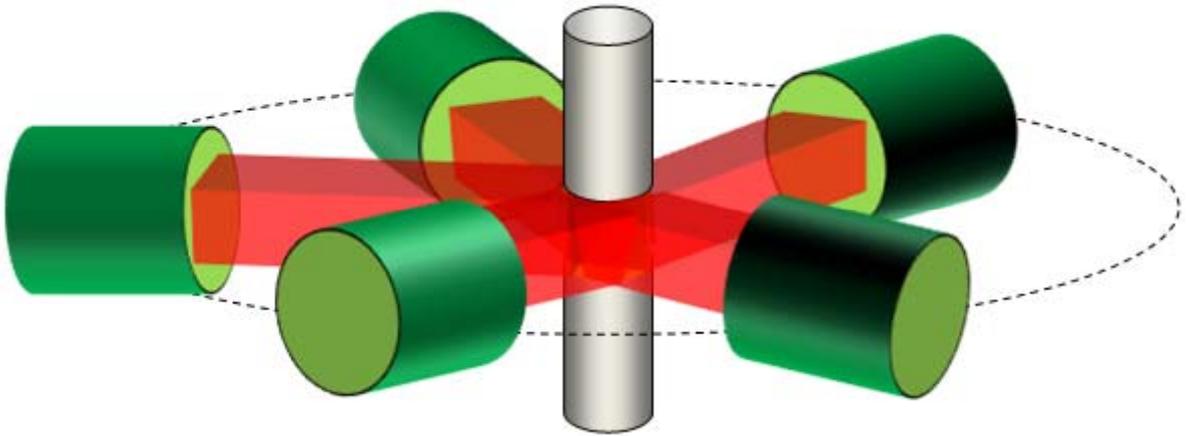

Fig. 2

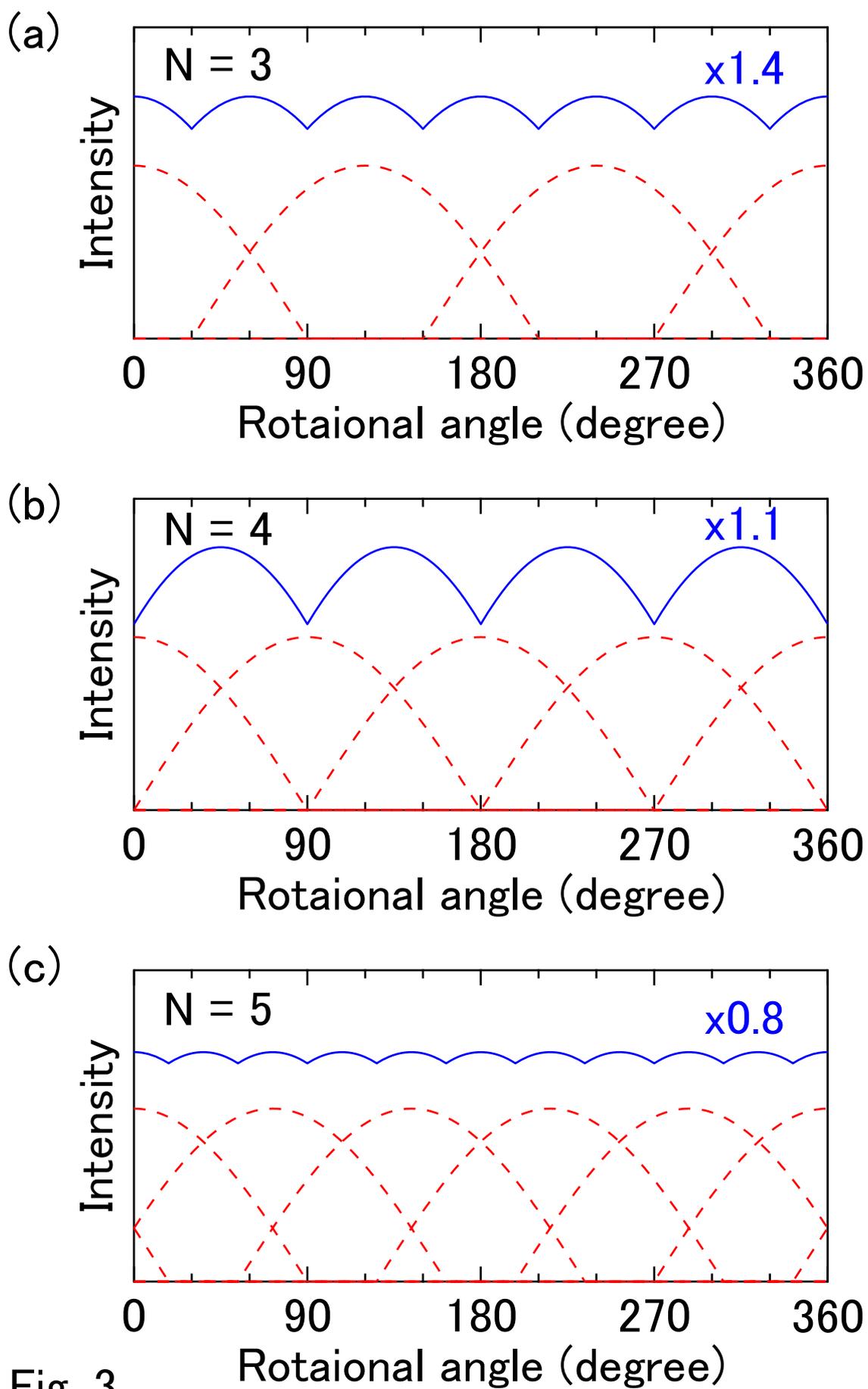

Fig. 3

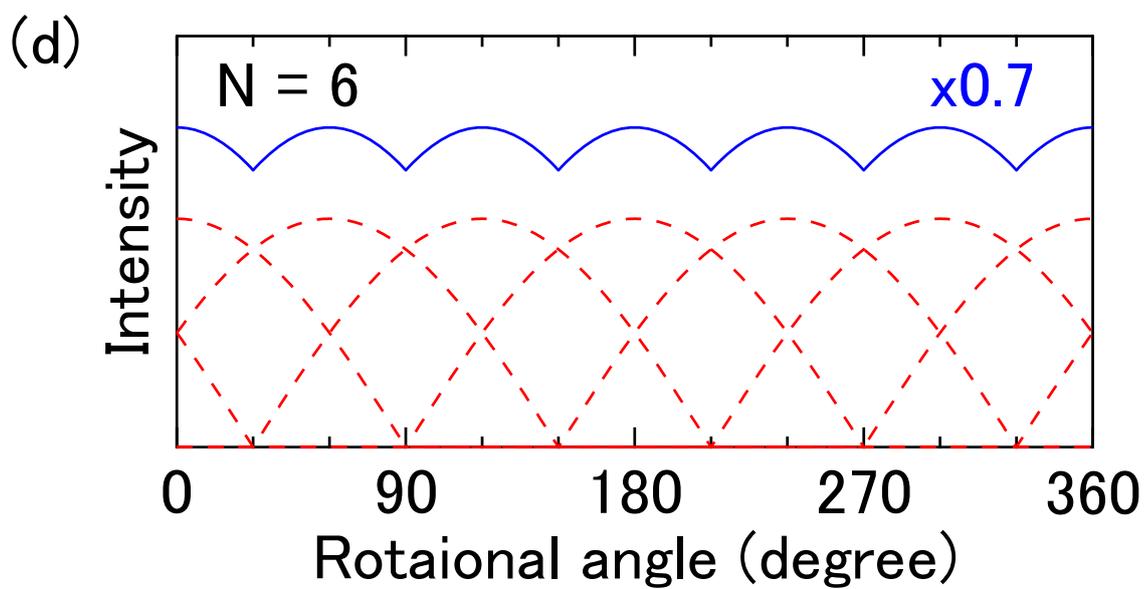

(d) N = 6, ×0.7

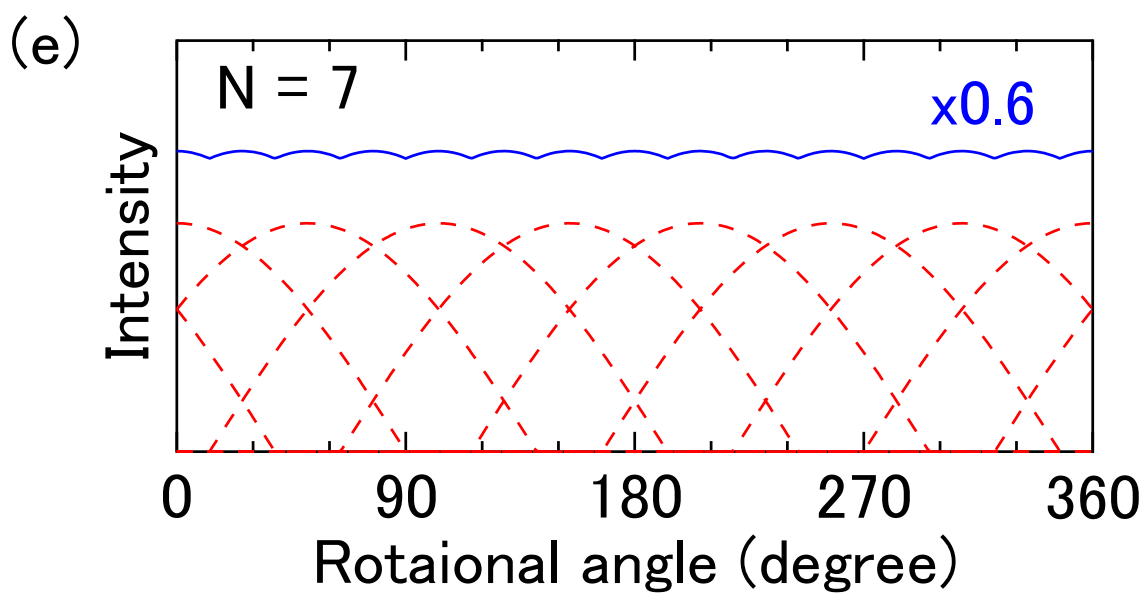

(e) N = 7, ×0.6

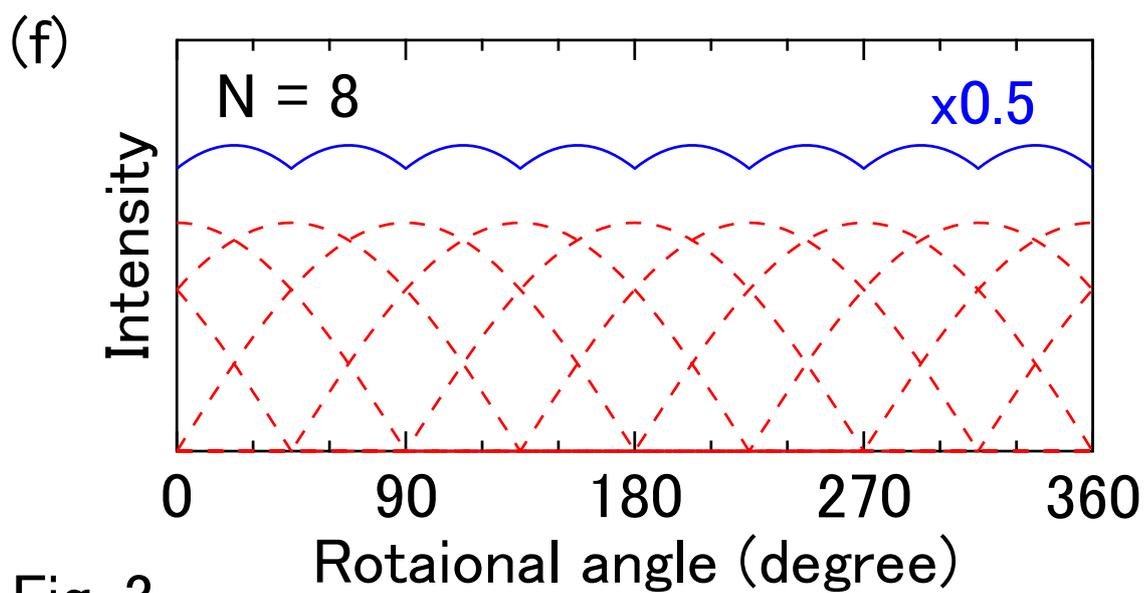

(f) N = 8, ×0.5

Fig. 3

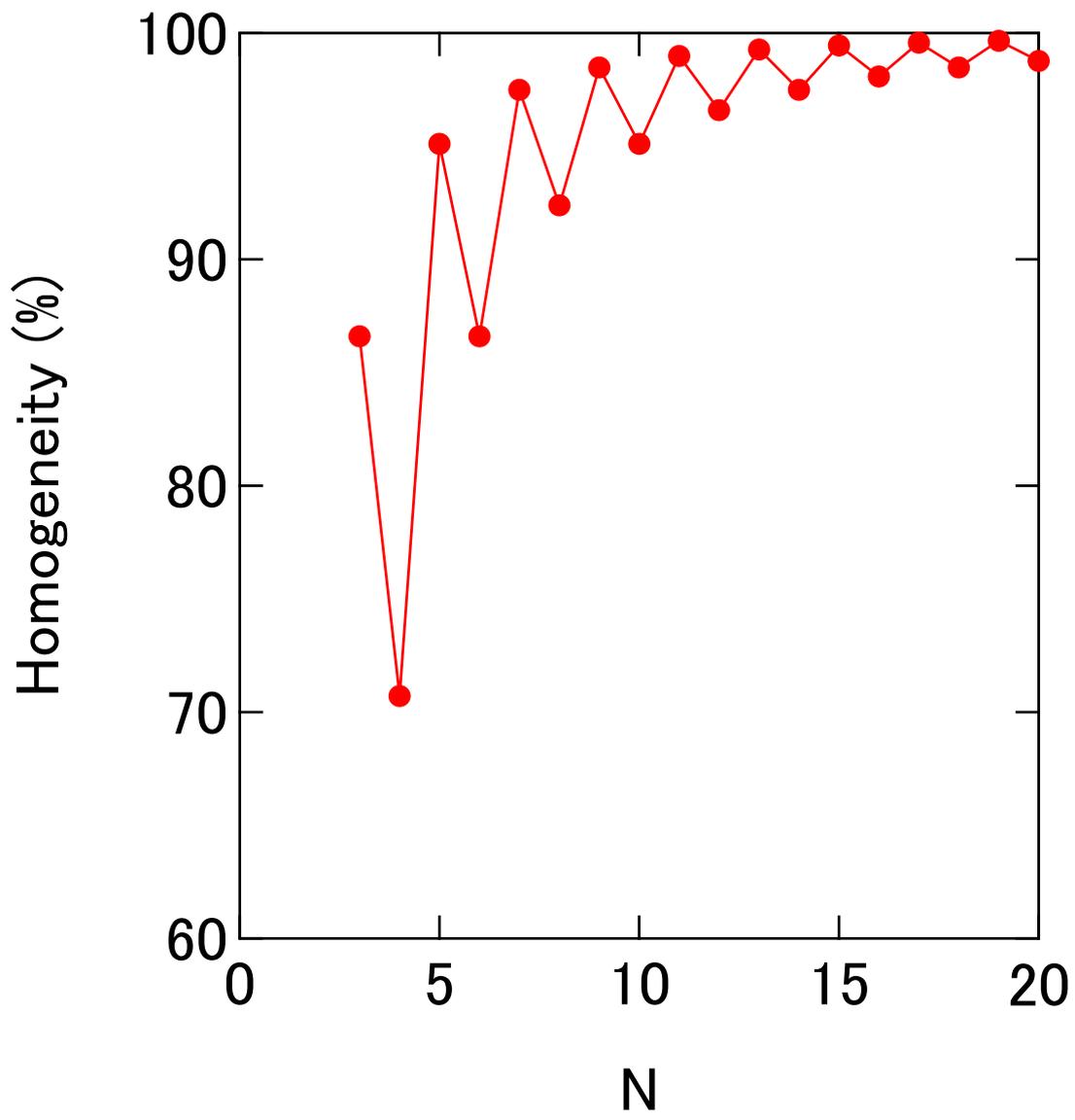

Fig. 4

(a)

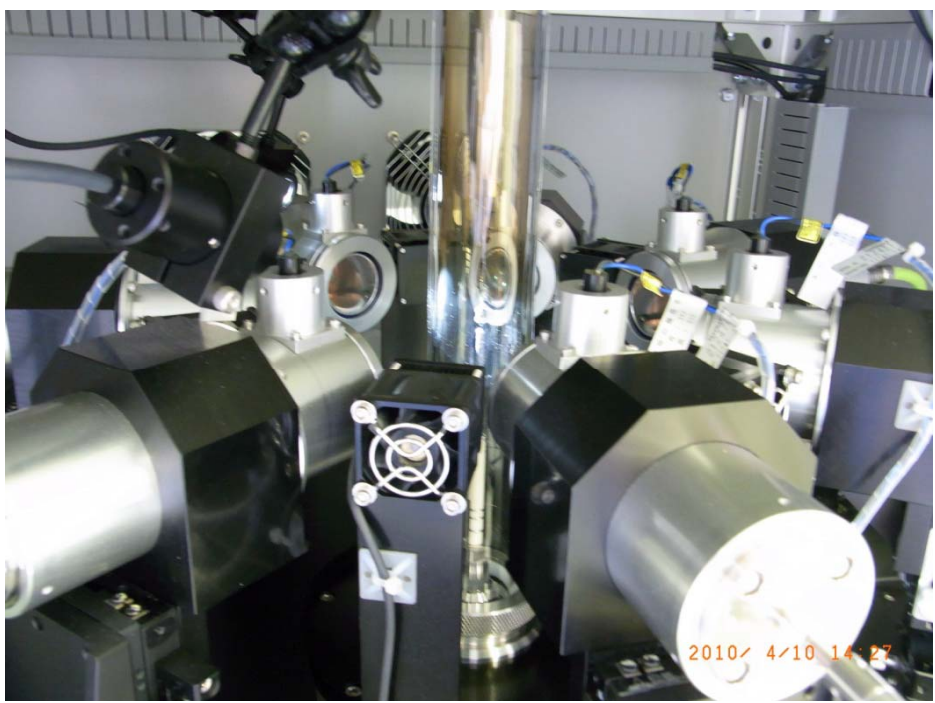

(b)

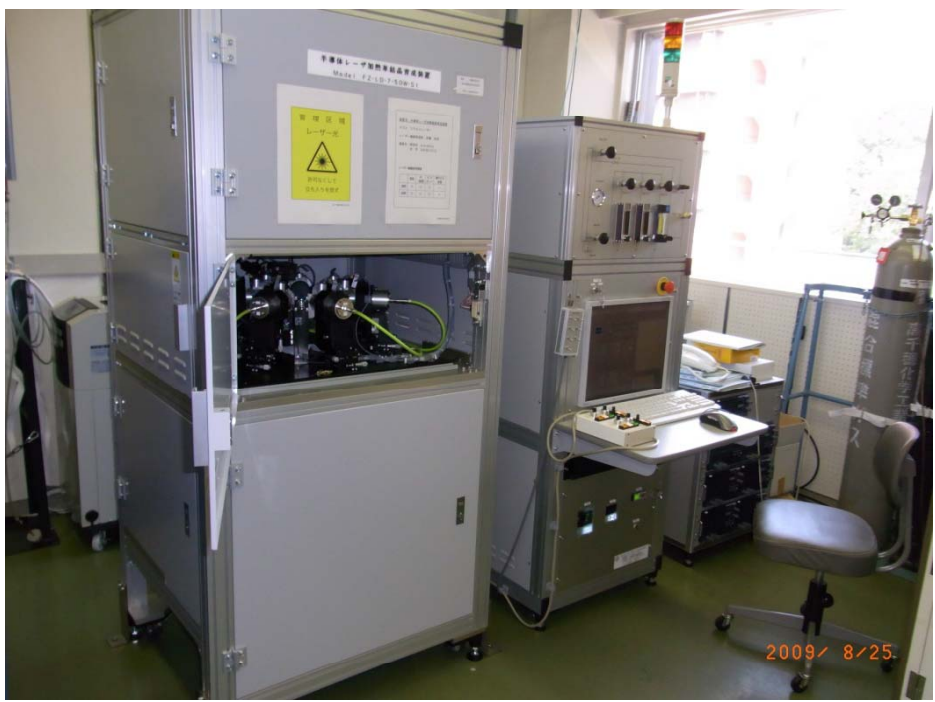

Fig. 5

(a) 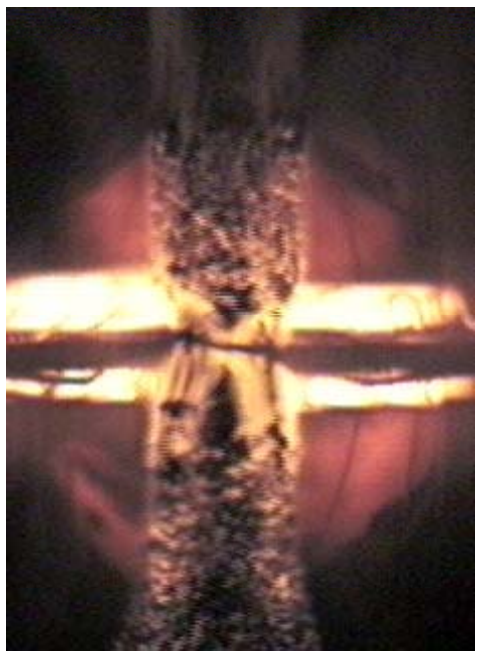

(b) 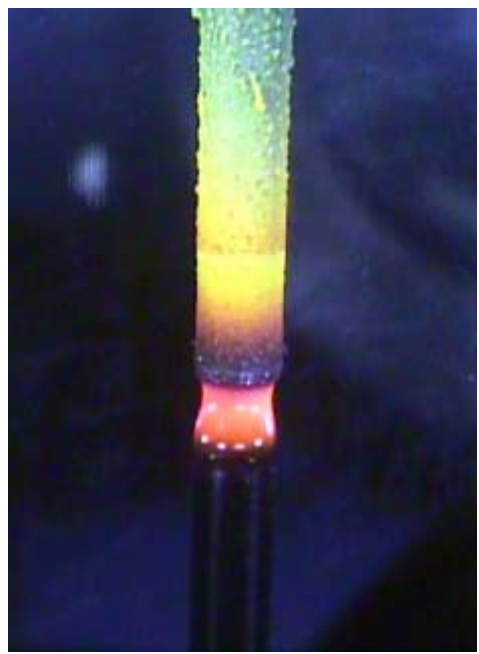

(c) 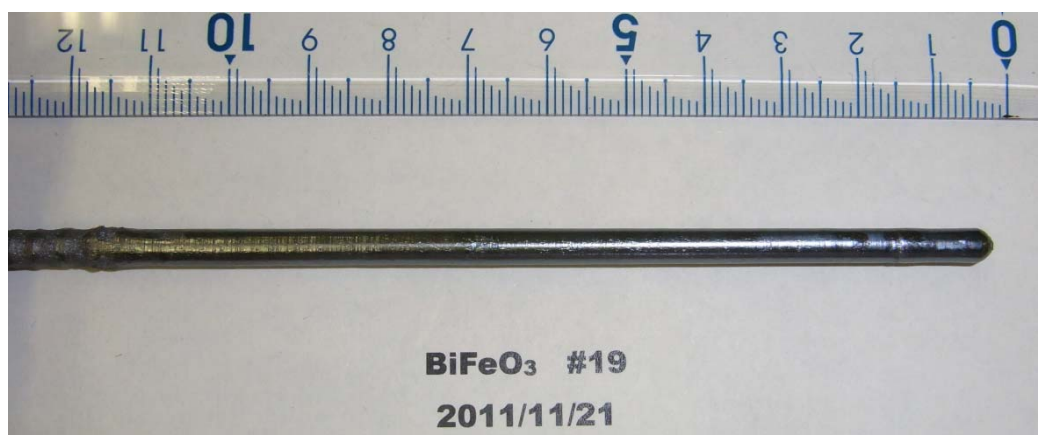

Fig. 6

(a)

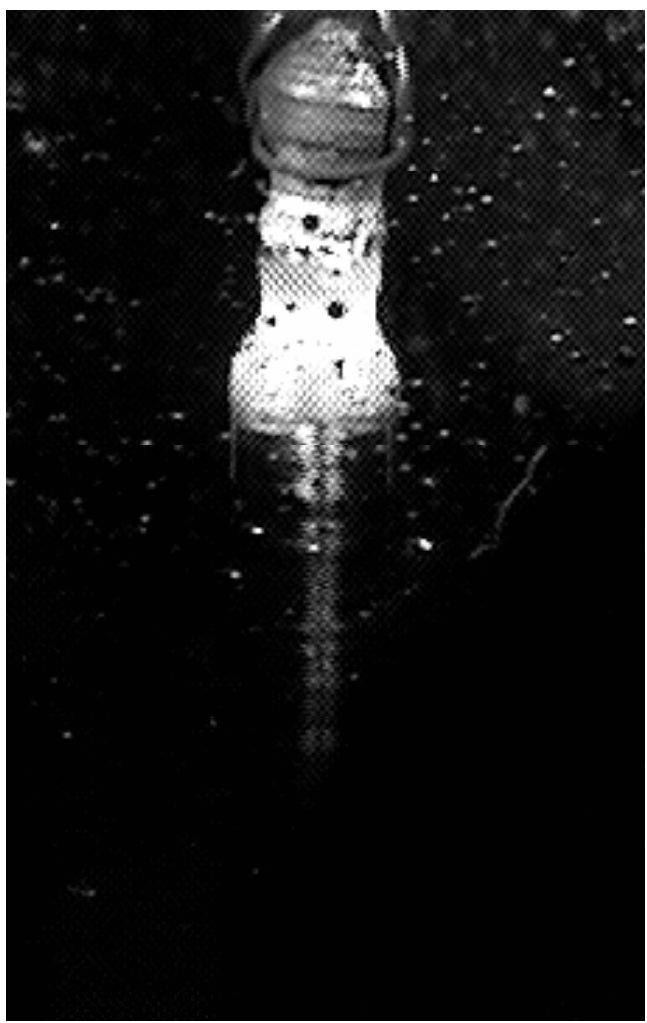

(b)

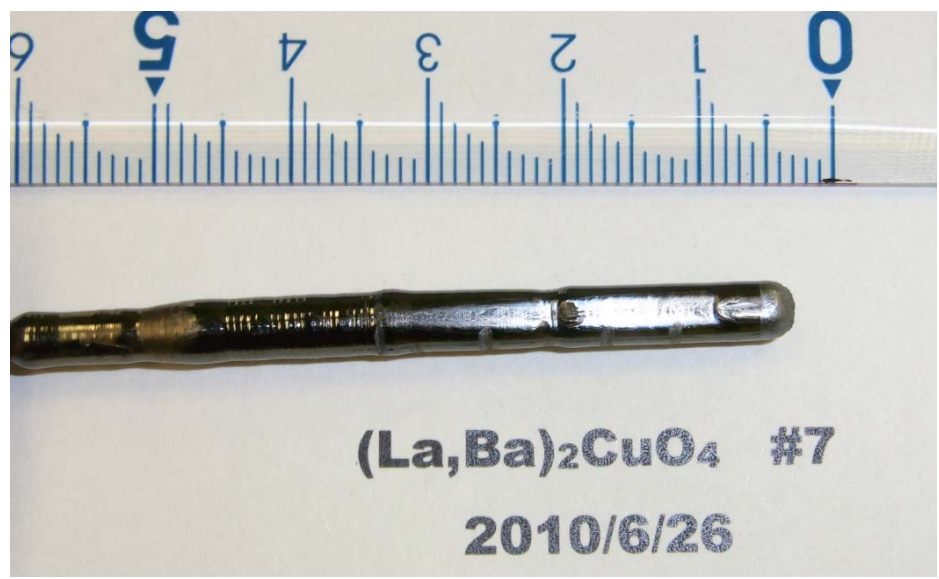

Fig. 7